\documentclass[twocolumn,showpacs,preprintnumbers,amsmath,amssymb]{revtex4}

\usepackage{graphicx}
\usepackage{dcolumn}
\usepackage{bm}

\begin{document}

\preprint{Taipex/AI-Econ/2007513}

\title{Network Topology of an Experimental Futures Exchange}

\author{Sun-Chong Wang}
\altaffiliation[Also at ]{Epigenetics Laboratory, Centre for Addiction and Mental Health, Toronto Ontario M5T 1R8, Canada}
\affiliation{
Institute of Systems Biology and Bioinformatics\\ National Central University,
Chungli 320 Taiwan}
\author{Jie-Jun Tseng, Chung-Ching Tai, Ke-Hung Lai, Wei-Shao Wu,
Shu-Heng Chen}
\affiliation{
AI-Econ Research Center and Department of Economics\\ National Chengchi University, Taipei 116 Taiwan}
\author{Sai-Ping Li}
\affiliation{%
Institute of Physics, Academia Sinica, Taipei 115 Taiwan
}

\date{\today}

\begin{abstract}
Many systems of different nature exhibit scale free behaviors. Economic 
systems with power law distribution in the wealth is one of the examples. 
To better understand the working behind the complexity, we undertook an
empirical study measuring the interactions between market participants.
A Web server was setup to administer the exchange of
futures contracts whose liquidation prices were coupled to event outcomes.
After free registration, participants started trading to compete for
the money prizes upon maturity of the futures contracts at the end of
the experiment. The evolving `cash' flow
network was reconstructed from the transactions between players.
We show that the network topology is hierarchical, disassortative and
scale-free with a power law exponent of
1.02$\pm$0.09 in the degree distribution.
The small-world property emerged early in the experiment while the number
of participants was still small.
We also show power law distributions of the net incomes and
inter-transaction time intervals. Big winners and losers are associated with
high degree, high betweenness centrality,
low clustering coefficient and low degree-correlation. We identify communities
in the network as groups of the like-minded. The distribution of the
community sizes is shown to be power-law distributed with an exponent of 
1.19$\pm$0.16.

\end{abstract}

\maketitle

\section{\label{sec:level1}Introduction}
Many complex systems exhibit distributions of observables that are not
characterized by a single scale.
Examples include net wealth,
earthquake magnitudes and
gene expression \cite{ueda04}. Heterogeneity in system constituents
and/or in the interactions among them
might underlie the complexity. Continuing
advances in information technology
have facilitated acquisition and analysis of sheer amounts of data, unraveling
the interacting networks of different kinds ranging from
the transportation network of airlines in
technology \cite{guimera05},
collaboration networks of
scientists in sociology \cite{newman01}
and binding networks of proteins in biology \cite{uetz00}.
Network topologies evolve to fulfill system requirements. Studies of networked
systems thus help better understand complex systems.
Among the encouraging examples are
the jamlessness of scale-free communication networks \cite{toroczkai04},
short separation of small-world acquaintance networks \cite{watts98} and
robustness against random mutations of scale-free biological
networks \cite{albert00}.
Further applications
of network analysis involve demarcation between social and nonsocial
networks by an attribute that measures the correlation between the degrees of
interacting nodes \cite{newman03}.
The finding of hierarchical structures in metabolic networks also has
implications for functional categorization of metabolites \cite{ravasz02}.

Financial markets, consisting of such heterogeneous agents as investors,
hedgers and arbitragers, show stylized
distributions of returns and wealth \cite{mandelbrot63,bouchaud97}.
Intrigued by the universal behavior, physicists
have applied the methodologies of nonequilibrium
statistical mechanics to elucidating the
mechanisms underlying the complexity \cite{stanley00}.
Examples include critical phenomenon \cite{stanley02}
and self-organized
criticality \cite{scheinkman94} modeling of economic systems.

In line with the network approach to technological, social
and biological complex systems,
we designed an experimental market, recording every transaction
between pairs of participants during 
the experiment. Transactions (edges) hold information on the flow of
assets from sellers to buyers (nodes). 
Characterization of the evolving topology of the resulting network
helps shed light on the emergence of complexity in financial markets.
The unique feature of our experiment is that no parallel 
can be easily undertaken in the real market. 
We describe the experimental settings and
market rules in Sec. II, followed by characterization of the network by 
mean shortest path lengths and degree and wealth distributions of Sec. III. 
Further analysis in Sec. IV unravels subtle network structures 
including hierarchy, 
dissortativity and community.
We argue that an integrated model of financial markets should accommodate
the results of our empirical study.

\section{Experiment}
A 24-hour exchange market was established
on the Web,
accepting bid and ask orders from registered players via the Internet
\cite{wang04,wang06a,wang06b}.
Upon registration, which was anonymous and free,
an account with 30,000 units of fictitious money
was allocated to the player on the exchange server. The futures contracts
that our
market issued were tied to the candidates running for the Taipei
mayoral election which took place on December 9,
2006\footnote{Contract specifications and trading rules were announced at http:// socioecono .phys .sinica .edu.tw/ exchange/ announce}.
The liquidation price of each futures
contract was determined by the percentage of votes the candidate received on the
election day. Such an experiment was run continuously
for 30 days, ending on the election
day.
After the experiment, any contracts in the players' 
accounts were liquidated using
the official counts released by
the government. Money
prizes were awarded to the top ten winners
determined by the accumulated wealth in the players' accounts.
In a previous publication \cite{wang06b}, we
demonstrated that such a market, which drew typically 400 participants,
exhibited power-law distributions of price
changes, net wealth and inter-transaction times that are characteristic of
real world markets. Furthermore, predictions of the market have so far
been consistent with election outcomes.
In this paper, we examine the evolving network of `cash' flow recorded
along the experiment.

\begin{figure}
\includegraphics[height=5.0cm,width=6cm]{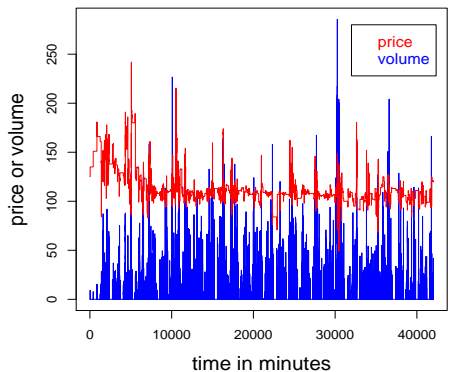}
\caption{\label{fig:epsart} (Color online) Price and volume time-series. The
time resolution is one minute. Prices (volumes) from multiple transactions
within one minute are averaged (summed).}
\end{figure}
\begin{figure}
\includegraphics[height=5.0cm,width=8cm]{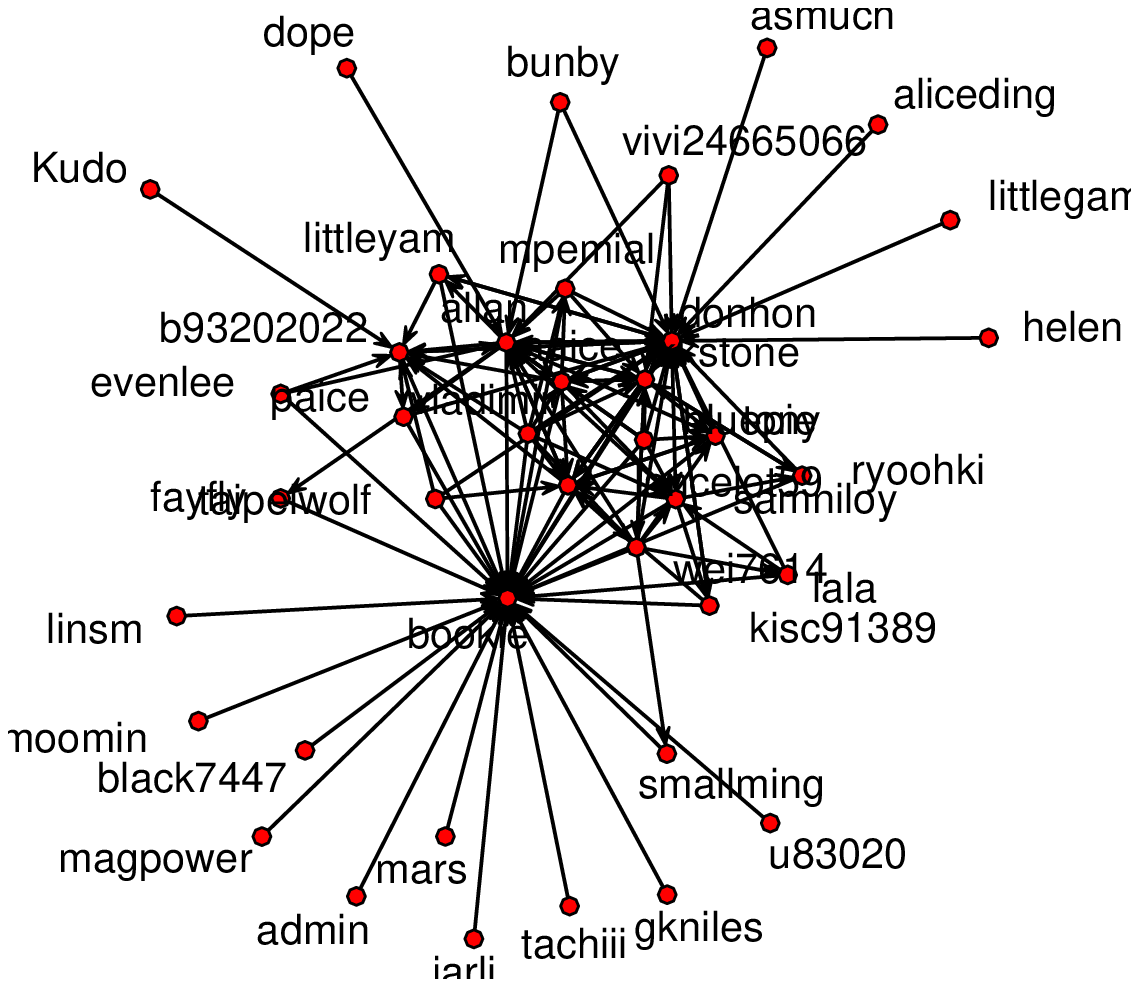}
\caption{\label{fig:epsart} The trading network on day 3 consisting of
40 interconnected nodes.
15 isolated nodes are not shown.}
\vspace{11pt}
\includegraphics[height=5.0cm,width=6cm]{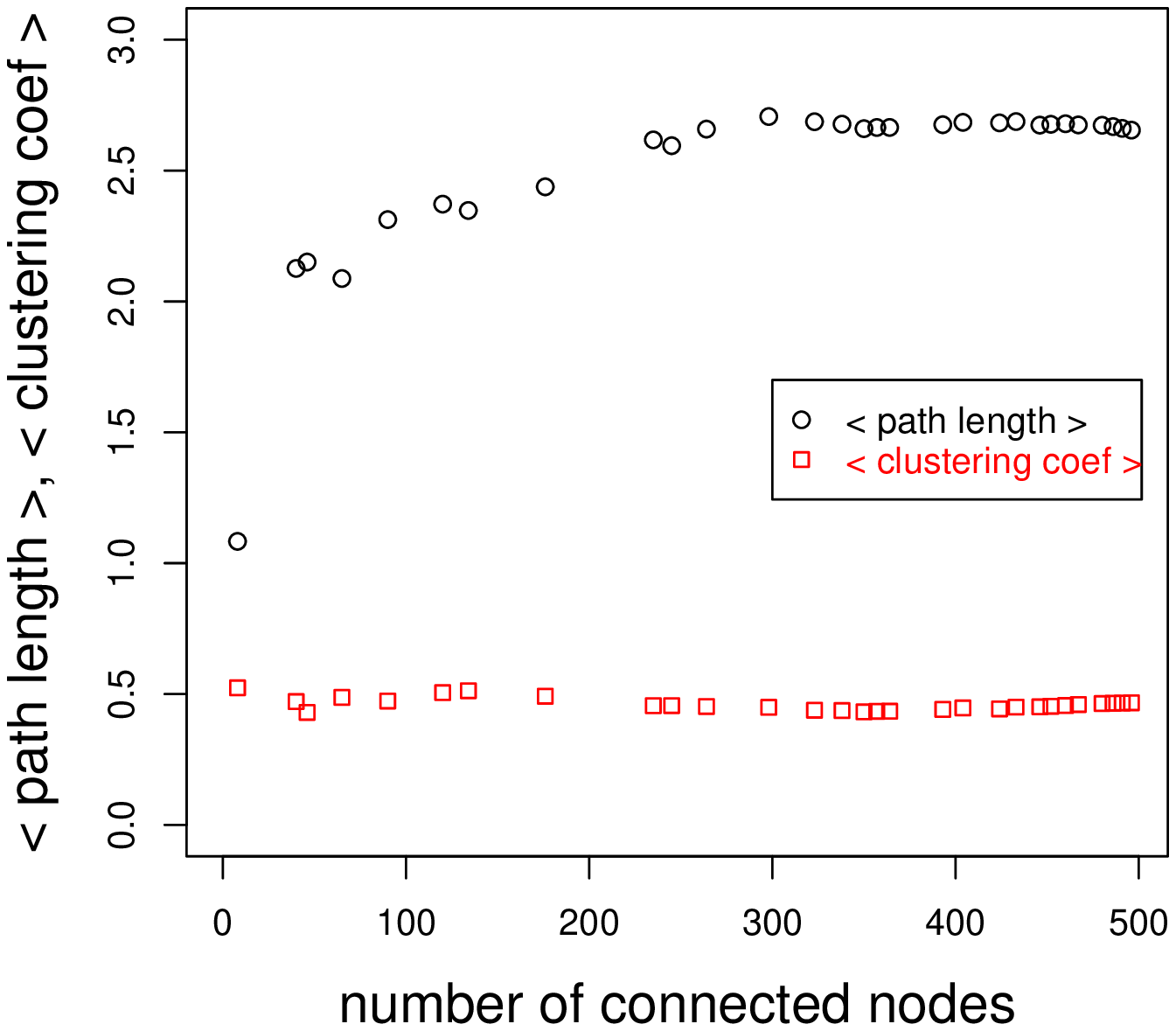}
\caption{\label{fig:epsart} (Color online)
Characteristic path length and clustering coefficient
as a function of network size.
}
\end{figure}

Five candidates ran for Taipei mayor. We included a sixth futures contract to
account for invalid ballots. The sum of the
six prices should be 100 if the players were rational or
the market was efficient. Figure 1 plots this
price and volume time series throughout the experiment. The
intermittence of price spikes may be attributable to
a multiplicative process
with additive noise which is known to yield power law distributions in
the fluctuations \cite{takayasu97,sornette98}. From
the number of time points where the trading
volumes are nonzero, we determine that the market was
active 12.7\% of the time. We advertised the experiment by constant posts
to the electronic
bulletin boards of the colleges throughout Taiwan during the experiment.
The number of registrants
increased roughly
monotonically with time, topping at 628 in the end of the experiment.
Trading orders submitted by players were stored in
the orderbooks on the server with the continuous double auction mechanism
for order matching and price finding. The number of successful
transactions in the experiment totaled 8,563.
Information on each of the transactions, including
price, volume, contract, buyer, seller and time, was recorded.

\section{Results and Analysis}
\subsection{Small World Cash Flow Networks}
When the bid order of player $i$ was matched with the ask order of player $j$
at a price $p$ and specified volume $v$ of a futures contract,
an amount of cash $p\times v$ flew from player $i$ to $j$. Every day,
the server output the cumulative
cash flow between any pair of players, from which we reconstructed
30 networks of cash flow, one for each day.
On average, 23$\pm$2\% of the nodes in the
networks were isolated,
corresponding to those who registered but had never traded with others.
Figure 2 shows the network on day 3. The day 1 network consists
only of three isolated nodes. 
The day 2 network has 11 nodes with 8 connected shown in
Fig~10. 
We exclude the isolated nodes in the following analysis.
The average number of (undirected) edges per node $<$$k$$>$ in the
network increased with day to about 6 within the first 10 days
and saturated at around 8 in the final days.
Inspection of the networks such as Fig.~2
by eyes identifies hubs which usually
confer the small world property. To confirm the property, we
calculate the characteristic path length of a network which is the average of
the smallest numbers of edges between pairs of nodes.
The slow increase of the characteristic path length with network size
in Fig.~3,
together with the high clustering coefficients (also shown
in Fig.~3 but to be elaborated later), demonstrates the small-worldness of the
cash flow networks.
The emergence
of the small world
property at early onset of the experiment suggests
a low quorum for such a market to function efficiently in terms of opinion
exchange.

\begin{figure}
\includegraphics[height=5.0cm,width=6cm]{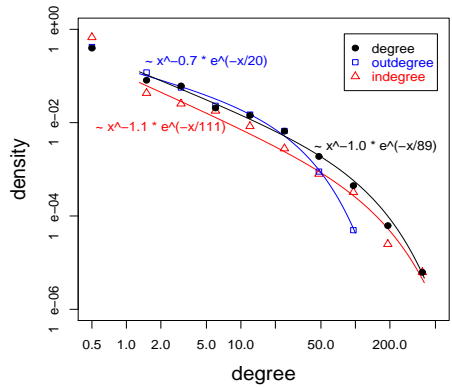}
\caption{\label{fig:epsart} (Color online)
Degree distribution of the cash flow network on the last day.
Black dots are from undirected
edges. Solid lines are least-squared fits to the data.}
\end{figure}
\begin{figure}[t]
\includegraphics[height=5.0cm,width=6cm]{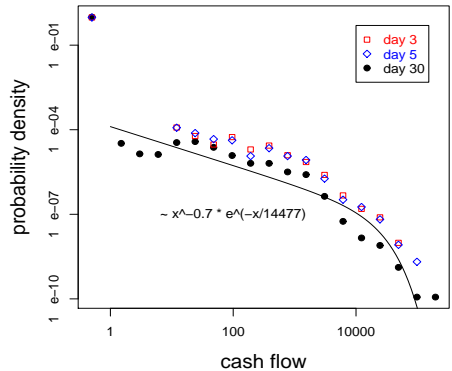}
\caption{\label{fig:epsart} (Color online) Probability densities
of the cumulative cash flow on the edges of the
networks on days 3, 5 and 30.
Solid lines are fits to the data.}
\end{figure}

\subsection{Degree Distribution}
The degree distribution $p(k)$ of a network gives the probability of a
randomly chosen
node to have $k$ edges. A power-law decay of $p(k)$ with $k$ indicates
excessive presence of hubs in the network.
To get the density distribution, we employ bin sizes that are even in the
logarithmic scale, with a binsize ratio of 2.
Figure 4 shows the distributions
of in-degrees, out-degrees and undirected degrees of the
cash flow network in the end of the experiment.
We found that the degree distributions are well described by a power-law
with exponential cut-off,
\begin{equation}
p(k)\sim k^{-\gamma}e^{-k/k_c}.
\end{equation}
The exponents
for the in-, out- and undirected degree are found to be
$\gamma_{in}=1.10\pm0.15$, $\gamma_{out}=0.66\pm0.13$ and $\gamma=1.02\pm0.09$,
respectively. The power laws in Fig.~4 show that the cash flow network from 
our experiment is scale free. Note however that
the small values of the exponents are in contrast
to those of other real world networks
found typically in the range $2 < \gamma < 3$.
Hubs play a pivotal role in opinion/information collection and dissemination.
If consensus is to be reached independent of the network size,
we would expect
a wide range of node degrees (corresponding to small $\gamma$'s) \cite{sood05}.
As we found that, despite transient spikes, the prices of individual
contracts were stationary as new players joined throughout the
experiment\footnote{Time-series plots of the individual contract prices are available at http:// socioecono .phys .sinica .edu.tw / exchange/ D/ TWMayors 06/ tw\_taipei06-p.jpg},
this property may explain the
small exponents.
The exponential cut-off could be due to such finite
size effects as the limited time frame and trade activity 
of the experiment.

\subsection{Weighted Networks and Wealth Distribution}
Flow of cash between players accumulated
as time went on.
We assign the cumulative flow of cash to the edge. The networks are
therefore weighted. The frequency distributions of the weights in Fig.~5
show that the weights are power-law distributed with an exponent of
0.69$\pm$0.11. Furthermore, the power-law weights behavior 
emerged in early stages of
the experiment as seen
from the distributions accumulated up to days 3 and 5 in Fig.~5.

We sum the weights on the directed edges pointing to (leaving from)
a node to obtain the
income (spending) of the node.
The incomes (and spending)
of the
nodes having the same degree are then averaged. A plot of the averaged
income versus degree is interesting in that it tells if high in-degree
players tend
to have high incomes. We found that the income and spending
increase with the in- and out-degree in a
power law fashion,
\begin{equation}
\begin{array}{c}
<\!{\rm incomes}\!>\, \sim k_{in}^{1.26\pm0.05}\\
\\
<\!{\rm spending}\!>\, \sim k_{out}^{1.02\pm0.07}.
\end{array}
\end{equation}
The relations indicate that those who managed to get
more buyers (sellers) cashed in
(spent) more.
We found no simple functional form relating
the in-
and out-degrees of the 496 active players up to the last day of the experiment
because of the divergence in the
scatter plot. However the nonparametric Spearman's rank
correlation
coefficient between the $k_{in}$ and $k_{out}$ is as high as 0.73.
The correlation indicates that those who cashed in more tended to spent more.
The same conclusion is reached if we symmetrize the cash flow matrix.
That is we replace $w_{ij}$=$A$ and $w_{ji}$=$B$ with
$w_{ij}$=$w_{ji}$={\it A}+{\it B}
where $w_{ij}$=$A$ means an amount of $A$ had flown from
player $i$ to player $j$ since the experiment began. {\it A}+{\it B}~
is then the traded amount between the two
players.
Using the symmetrized, weighted cash flow matrix,
we found a power law similar to Eq.~(2):
$<$traded$>$ $\sim k^{1.15\pm0.04}$
where $k$ is undirected degree.

To get the distribution of net incomes, we subtract the spending from the income
of each player. The probability densities in Fig.~6 show power law behavior of
the incomes, spending, earnings (positive net incomes) and losses
(negative net incomes), reminiscent of the Pareto distribution.
The exponent of the earnings is 0.99$\pm$0.04.

\begin{figure}
\includegraphics[height=5.0cm,width=6cm]{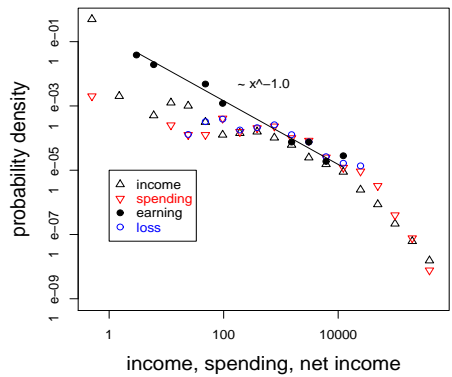}
\caption{\label{fig:epsart} (Color online)
Probability density distributions of the
players' incomes, spending and net incomes on the last day.
Solid line is fit to the data.}
\end{figure}
\begin{figure}
\includegraphics[height=5.0cm,width=6cm]{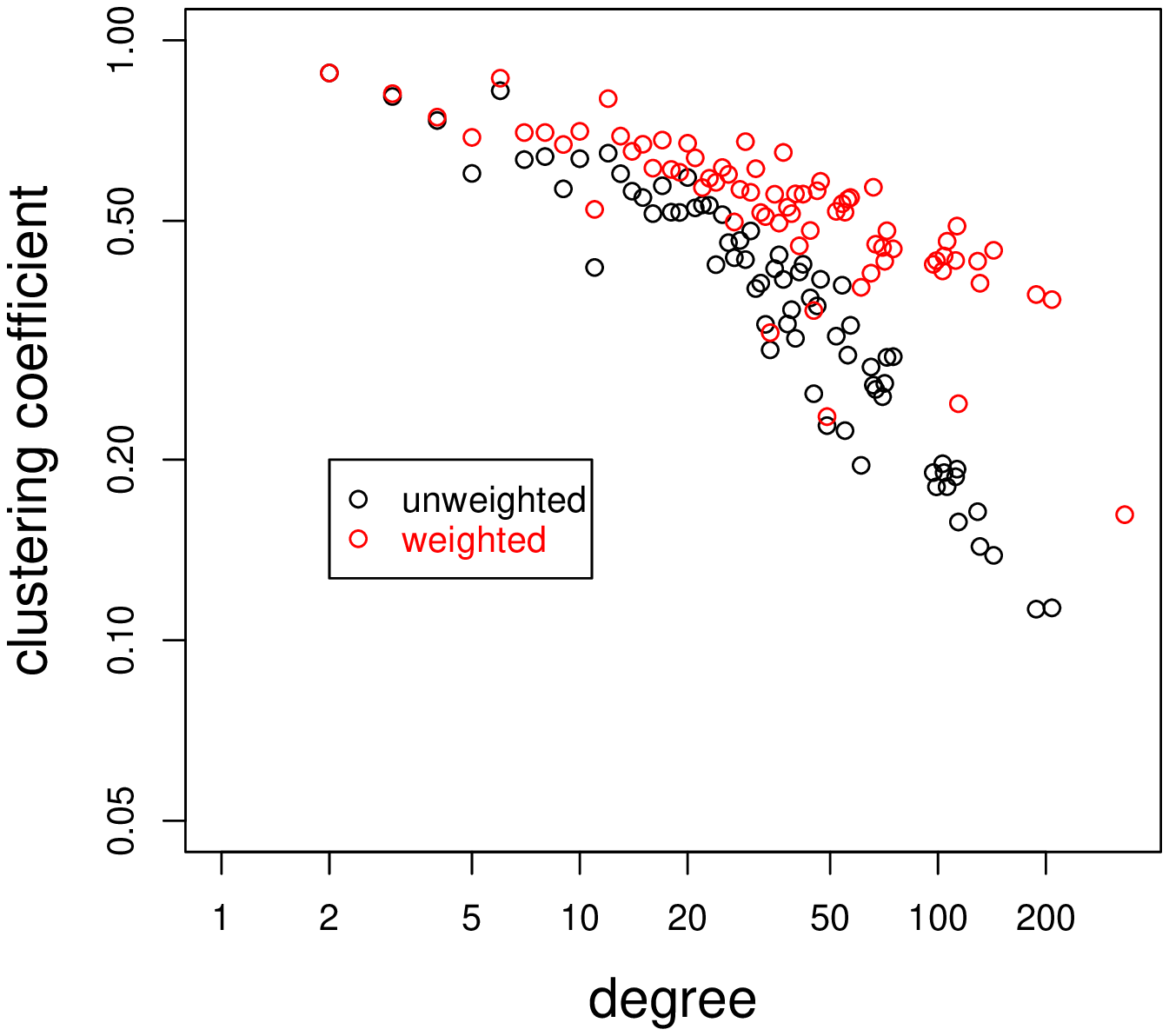}
\caption{\label{fig:epsart} (Color online)
Decrease in the clustering coefficient with degree.
The clustering coefficients are calculated from the cash flow network on
day 30.}
\vspace{11pt}
\includegraphics[height=5.0cm,width=6cm]{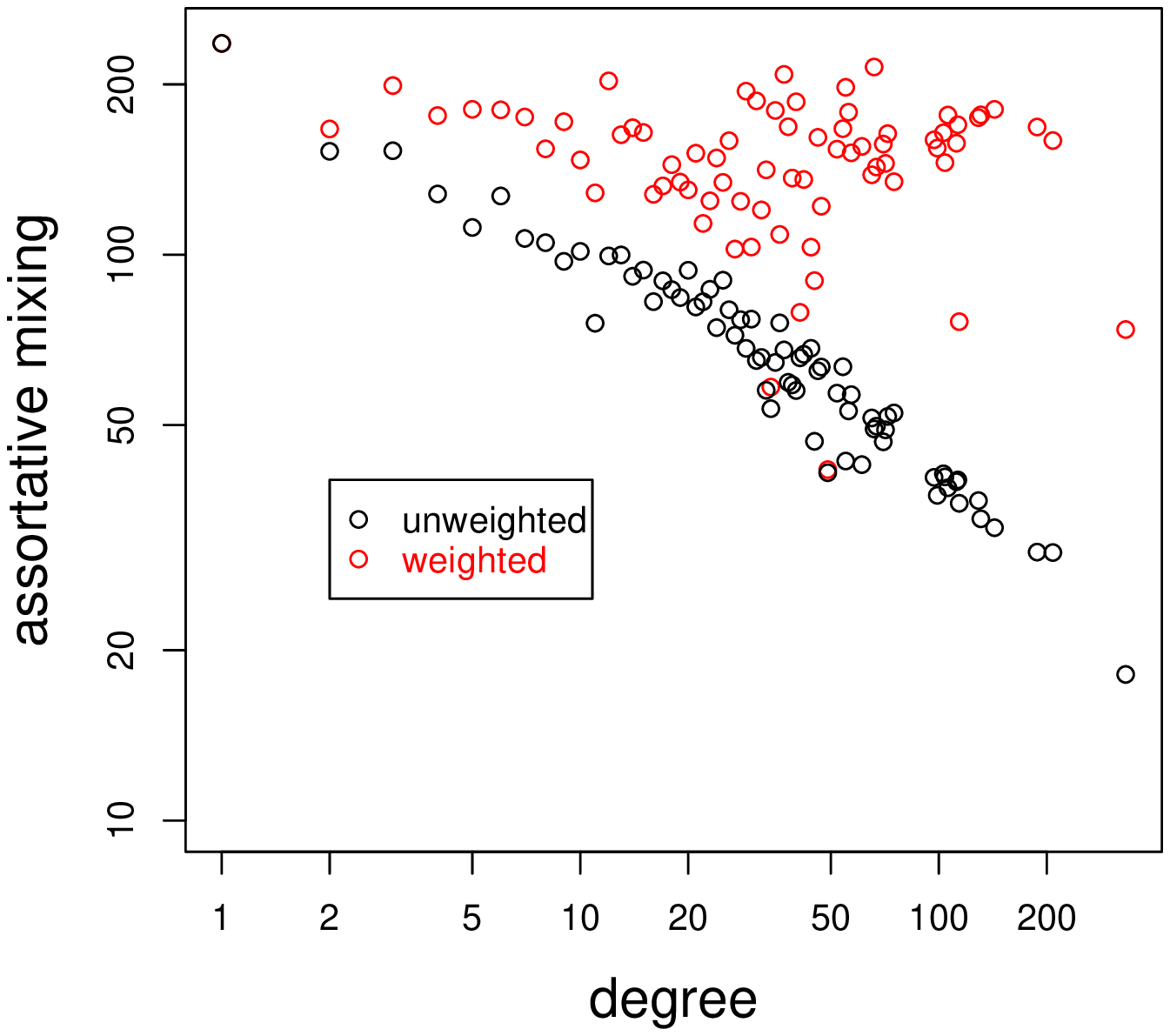}
\caption{\label{fig:epsart} (Color online)
Decrease in assortative mixing with degree. The assortative mixings are
calculated from the cash flow network on the last day of the experiment.}
\end{figure}
\begin{figure}
\includegraphics[height=5.0cm,width=6cm]{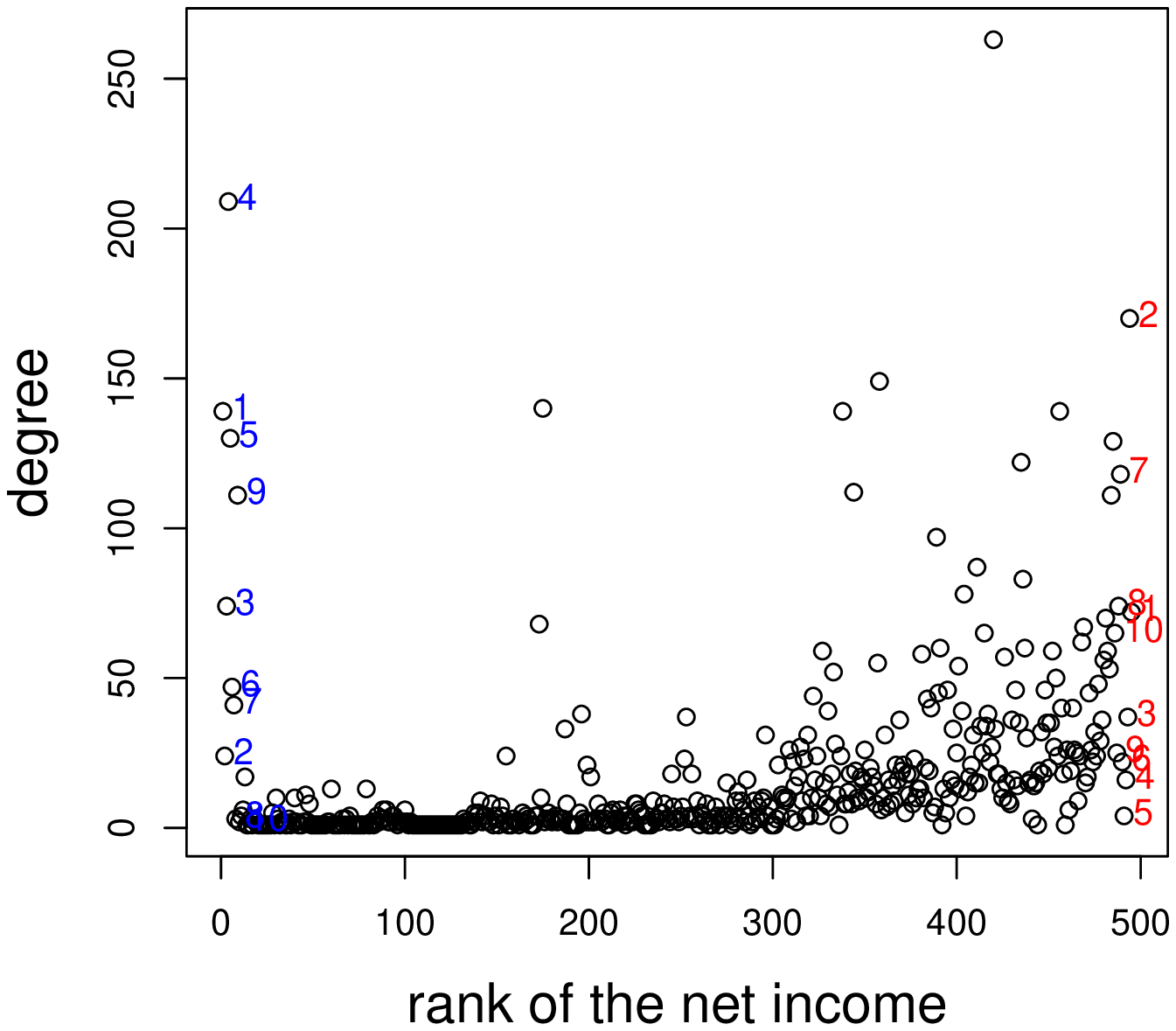}
\caption{\label{fig:epsart} (Color online)
Player's degree versus rank. Top ten winners are marked with blue
labels. The bottom 10 players (i.e. losers) are marked with red labels.}
\vspace{11pt}
\includegraphics[height=5.0cm,width=6cm]{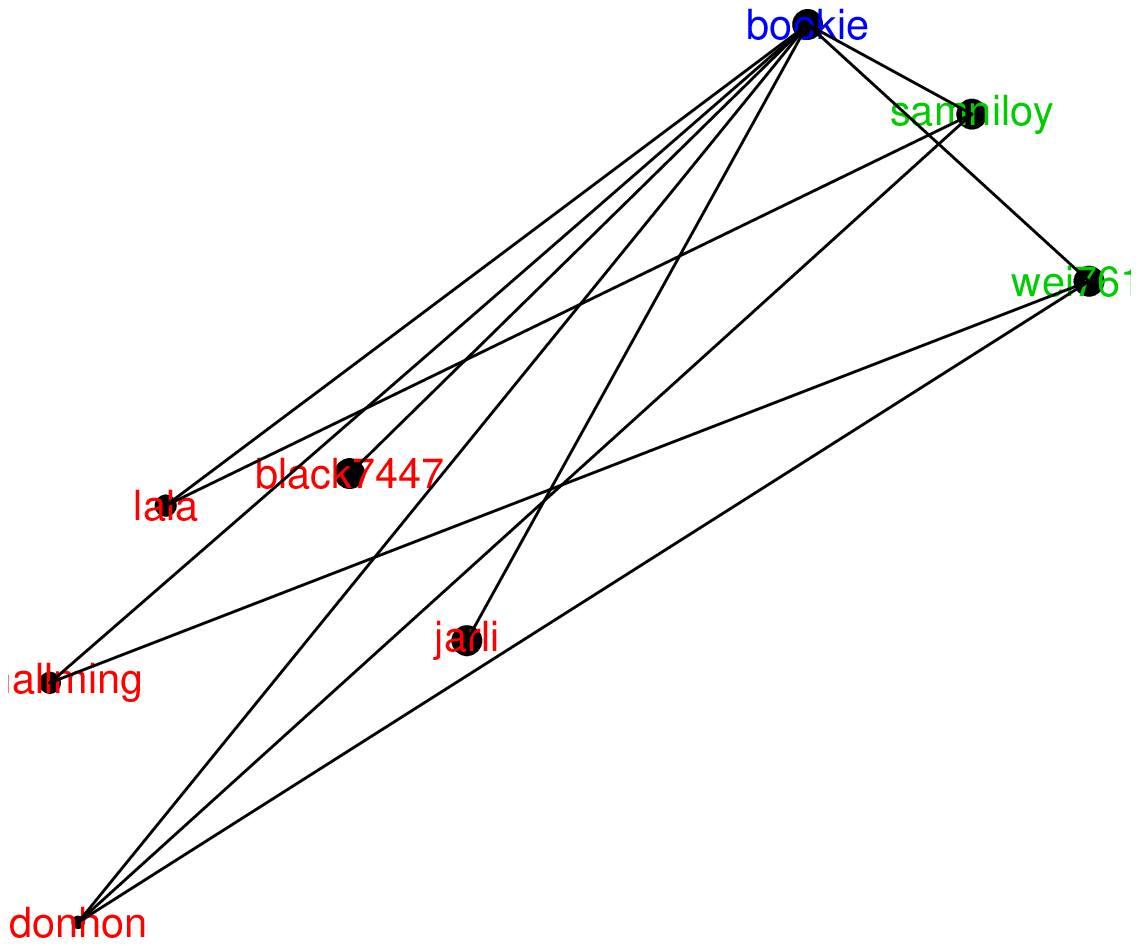}
\caption{\label{fig:epsart} (Color online) Communities in the cash flow 
network on day 2. Labels of the same color form a community which, by our
construction, contains higher-than-average density of links between communities
and lower-than-average density of links within communities.}
\end{figure}
\begin{figure}
\includegraphics[height=5.0cm,width=6cm]{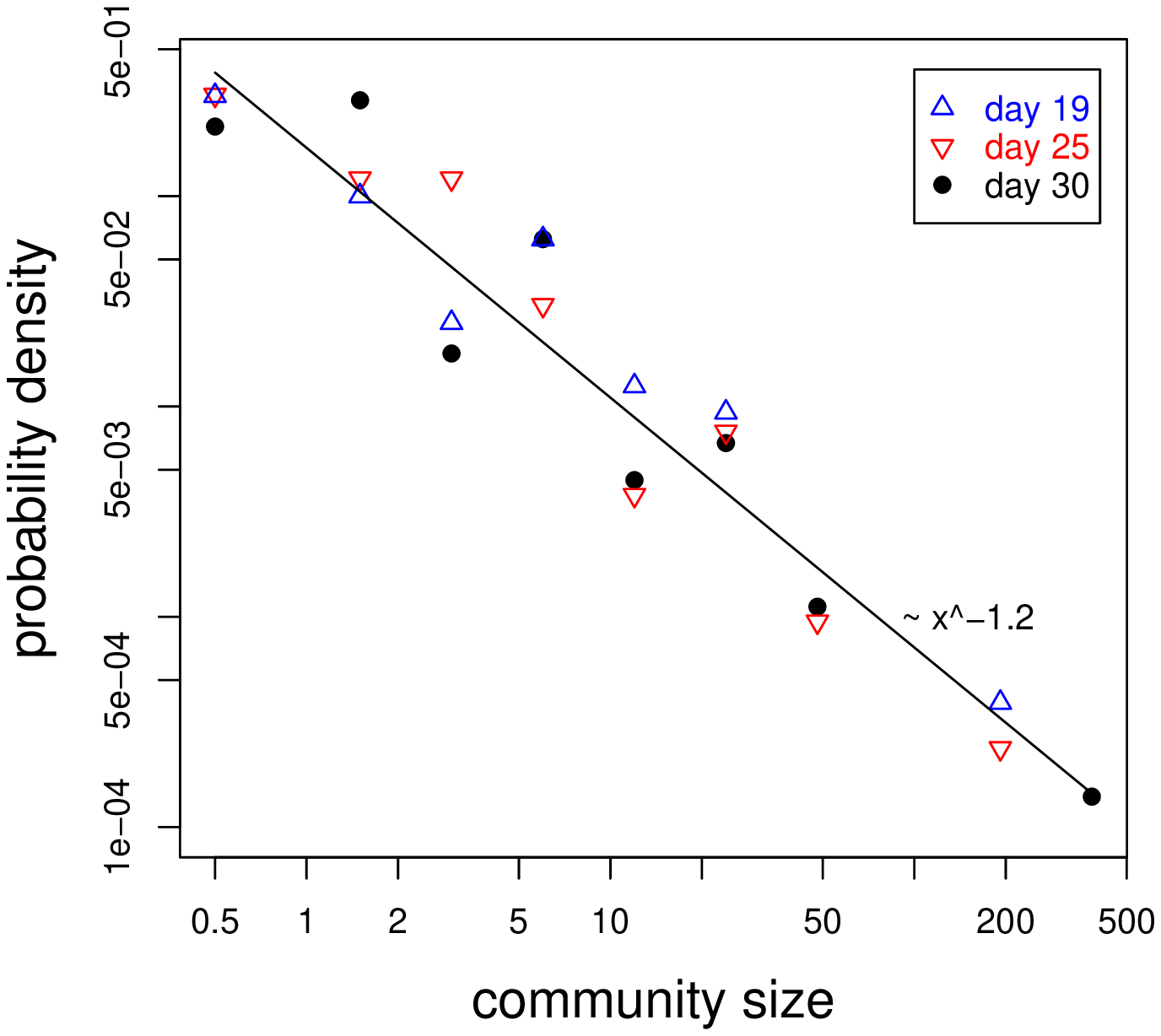}
\caption{\label{fig:epsart} (Color online)
Distributions of the community sizes from the cash flow networks 
on three 
terminal days. Solid line is fit to the distribution on day 30,
providing the exponent.
}
\vspace{11pt}
\includegraphics[height=5.0cm,width=6cm]{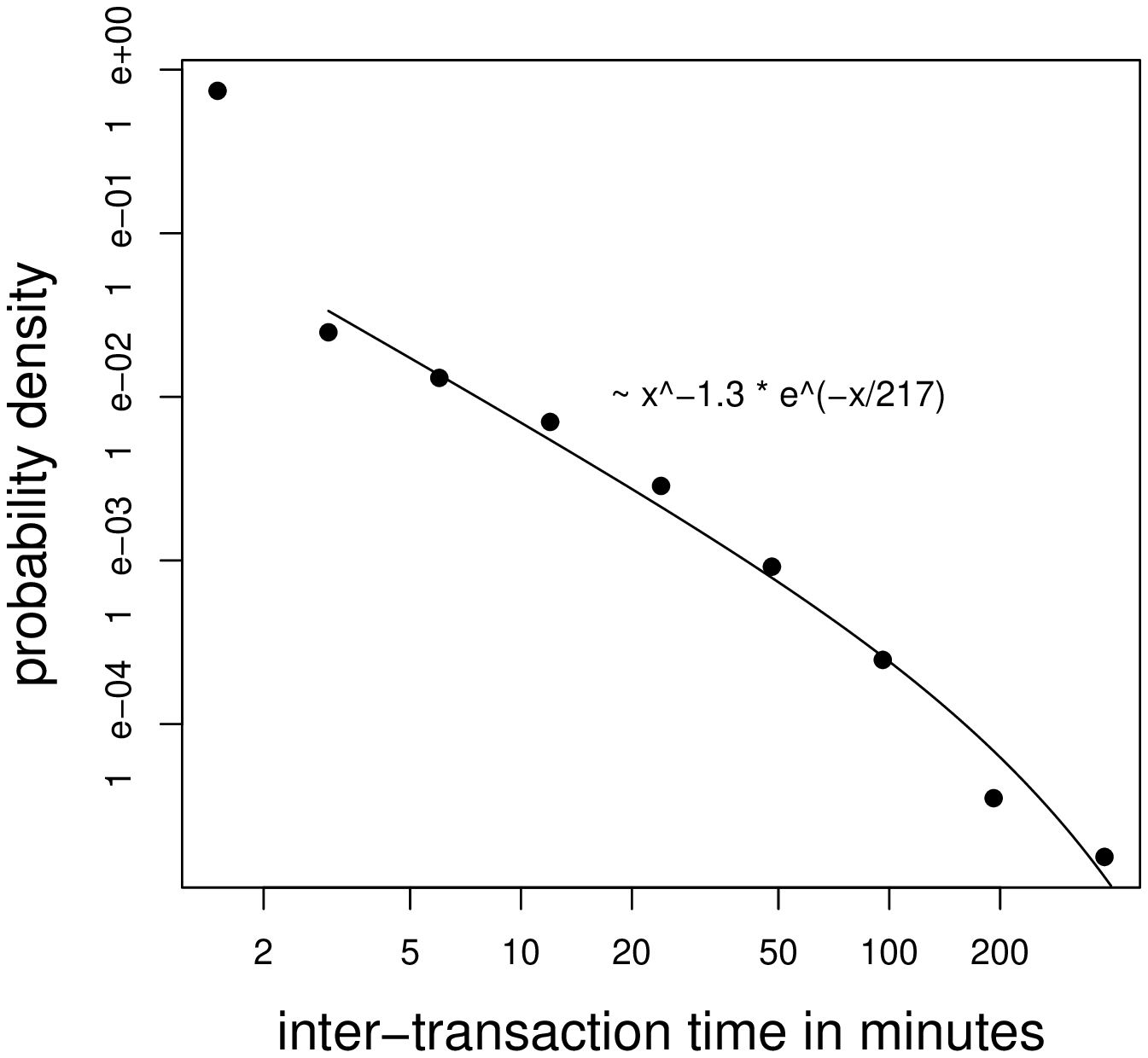}
\caption{\label{fig:epsart} Power law decay of the
time intervals between successive
transactions. Solid line is fit to the data.}
\end{figure}

\section{Discussion}
We have reconstructed the evolving network of `cash' flow between the players
who participated in the trading experiment to compete for real money prizes.
We showed that the cash flow network was scale free with small world properties
that emerged very early in the experiment.
The distributions of the weights (incomes
and spending) are also power-law decaying.

\subsection{Hierarchical Structure}
To further characterize the cash flow network, we calculate the clustering
coefficient of a player $c_i$ which measures the propensity that her
trading partners had traded with one another.
We use symmetrized adjacency $a_{ij}$ and cash flow $w_{ij}$
matrices in the calculation of the weighted clustering
coefficient \cite{barrat04}
which takes into account the frequency
or amount of trades between the players,
\begin{equation}
c_i={1\over s_i(k_i-1)}\sum_{j,h}{w_{ij}+w_{ih}\over 2}a_{ij}a_{ih}a_{jh}
\end{equation}
where
\begin{equation}
s_i = \sum_j w_{ij}a_{ij}.
\end{equation}
The clustering coefficients of the
players having the same degrees are averaged to get $c(k)$.
The results in Fig.~7 show a power law decay of the average
clustering coefficients $c(k)$ with increasing degree,
suggesting a hierarchical architecture in the cash flow network
\cite{ravasz03}.
We also calculated the average clustering coefficient $<$$C$$>=(1/N)\sum_i c_i$
for each of the 30 networks
and found that the values stay rather constant independent of the network
size $N$ (cf Fig.~3), the averages being 0.46$\pm$0.02 and 0.52$\pm$0.03
for the unweighted (Fig.~3) and weighted versions of the coefficient.

\subsection{Disassortative Mixing, Betweenness and Rank}
We also calculate the assortative mixing $k_{{nn}_i}$ which measures the
similarity between player $i$ and her trading partners in terms of their degrees
\cite{barrat04},
\begin{equation}
k_{{nn}_i} = {1\over k_i}\sum_j a_{ij}w_{ij}k_j.
\end{equation}
An analysis parallel to Fig.~7 shows
a decaying degree-correlation with increasing degree,
indicating that the cash flow
network is disassortative. The dissortativity may reflect the competitive nature
of the market although the dissortativity becomes insignificant
considering the weights on the edges (Fig.~8).

Another quantity in network analysis is the betweenness
centrality of node $i$
defined as the number of shortest paths between two other nodes
passing through $i$ weighted by the inverse of the number of
redundancies \cite{freeman77}.
We found the mean betweenness centrality $b(k)$ is related to degree by
$b(k)\sim k^{2.35\pm0.08}$.

Most of the properties of the node, such as clustering coefficient and
betweenness centrality, can be referred to its degree. We
rank the players according to their net incomes and plot the degree against
rank in Fig.~9. The plot shows that high degree players reaped either
victory or debacle.
The high degree players tend to have low clustering coefficients as in
Fig.~7. A low clustering coefficient translates that, instead of
trading within a clique of partners, the player keeps
searching for new investment opportunities across cliques over the network.
Whether she wins or loses would then depend on her
adaptability to changing opinions.

\subsection{Power-law Distribution of Community Sizes}
In the context of our experiment, when the price of a futures contract
was considered too high (low), a sell (buy) order was placed.
An edge between two nodes in the cash flow network therefore indicates 
that the two
players disagreed to the pricing of the futures contract.
In other words, players with no  
edges linking them were those who thought alike. 
An algorithm to find communities in the players is thus to partition the cash
flow network so that the density of edges within communities are 
lower and that between communities are higher than average.
An example of such a division of the network is shown in Fig.~10 where 
it is clear that the within-community edges are minimized while the
between-community
edges are maximized. We applied the eigenvector-based 
partitioning algorithm of \cite{newman06} to the 30 
networks and found that the number of communities grew logarithmically with the
number $N$ of active players as $-17+7\log(N)$. Furthermore, 
the distribution of 
community sizes, shown in Fig.~11, is found power-law distributed with an
exponent of $1.19\pm 0.16$. Figure 11 shows three distributions from three
cash flow networks on day 19, 25 and 30. The day 19 and 25 networks 
have, respectively, 20 and 33 
communities, corresponding to two maximal deviations ($-5$ and $+7$)
from the above best fit prediction.
The distributions in the plot demonstrate that
they are power-law distributed. Moreover,
the largest communities encompass
$\sim$61\% of the players. 
 
\subsection{Distribution of Inter-transaction Time Intervals}

We find the time intervals between successive transactions from
the ticks in the volume time-series in Fig.~1. The distribution
of the inter-transaction times shown in Fig.~12 exhibits a truncated
power law distribution
with exponent 1.28$\pm$0.17, consistent with our previous
finding \cite{wang06b}.
This power law behavior, together with
that in the early-day cash flows of Fig.~5, may
suggest a contribution of human factors \cite{barabasi05} to
the origins of power laws.

In summary, in an effort to study financial markets through network 
approach, we performed an online experiment in the form of tournament.
We recorded the flow of
fictitious cash between the 496 registered, 
active participants throughout the 30-day course
of the experiment.
The topology of the resulting cash flow networks is found
nonrandom with a power-law distribution in the connectivity.  
The heterogeneity in the connectivity as well as weights emerged early in the 
experiment.
The distribution of net incomes in the end of the experiment is also power-law
distributed. Network analysis indicates that the cash flow network is 
hierarchical and disassortative. Communities in the network are 
defined and identified.
The distribution of community sizes is power-law distributed, so is the
distribution of inter-transaction time intervals.
Our experimental platform offers a unique chance of anatomizing such
complex systems as 
financial markets. A better understanding of the complexity calls for
models that account for the major findings in the present study.

\begin{acknowledgments}
The research was supported in part
by the National Science Council of Taiwan
(NSC \#95-2112-M-001-010 and NSC \#95-2415-H-004-002-MY3).
\end{acknowledgments}

\bibliography{taipex}

\end{document}